\documentclass[pra,twocolumn,amsmath,amssymb,superscriptaddress,
showpacs]{revtex4}
\usepackage{graphics}
\usepackage{epsfig}
\usepackage{bbm}
\usepackage[latin1]{inputenc} 

\newcommand{\be}{\begin{equation}}
\newcommand{\ee}{\end{equation}}
\newcommand{\eea}{\end{eqnarray}}
\newcommand{\bea}{\begin{eqnarray}}

\newcommand{\eins}{\openone}
\newcommand{\qed}{\ensuremath{\hfill \Box}}
\newcommand{\WW}{\ensuremath{\mathcal{W}}}
\newcommand{\NN}{\ensuremath{\mathcal{N}}}

\newcommand{\FF}{\ensuremath{\mathcal{F}}}

\newcommand{\fO}{\ensuremath{\mathfrak{O}}}

\newcommand{\ketbra}[1]{\ensuremath{| #1 \rangle \langle #1 |}}

\newcommand{\ket}[1]{\ensuremath{|#1\rangle}}

\newcommand{\bra}[1]{\ensuremath{\langle#1|}}

\newcommand{\kommentar}[1]{}

\renewcommand{\vr}{\ensuremath{\varrho}}

\newcommand{\forget}[1]{}

\begin{document}
\title{Separability criteria for genuine multiparticle entanglement}
\date{\today}
\begin{abstract}
We present a method to derive separability criteria for the different 
classes of multiparticle entanglement, especially genuine multiparticle 
entanglement. The resulting criteria are necessary and sufficient for 
certain families of states. This,  for example,  completely solves the problem of classifying $N$-qubit 
Greenberger-Horne-Zeilinger states mixed with white noise according to their separability and entanglement properties. 
Further, the criteria are superior to all  known entanglement criteria for many other families; also they allow 
the detection of bound entanglement. We next demonstrate that they are 
easily implementable in experiments and discuss applications to the 
decoherence of multiparticle entangled states.
\end{abstract}

\author{Otfried G\"uhne}
\affiliation{Institut f\"ur Quantenoptik und Quanteninformation,
~\"Osterreichische Akademie der Wissenschaften,
6020 Innsbruck, Austria,}
\affiliation{Institut f\"ur Theoretische Physik,
Universit\"at Innsbruck, Technikerstra{\ss}e 25,
6020 Innsbruck, Austria}
\author{Michael Seevinck}
\affiliation{Institute of History and Foundations of Science \& Institute for Theoretical Physics, 
Utrecht University,
P.O Box 80.010, 3508 TA Utrecht, The Netherlands
}
\affiliation{Center for Time, Philosophy Department, University of Sydney, Main Quad A14, NSW 2006, Australia (visiting address)}


\maketitle

\section{Introduction}  
Entanglement is relevant for many effects in quantum optics 
or condensed matter physics and its characterization is 
of eminent importance for studies in quantum information 
processing \cite{hororeview, review}. Concerning entanglement 
between two particles, many questions are still open, but 
there exist at least various criteria which can be used to test 
whether a given quantum state is entangled or separable. 
For more than two particles, however, the situation is 
significantly more complicated, as several inequivalent 
classes of multiparticle entanglement exist and it is 
difficult to decide to which class a given state belongs. 
Entanglement witnesses and Bell inequalities can sometimes 
distinguish between the different classes \cite{review,bell}. 
However, it would be desirable to have useful criteria 
which allow to detect the different classes of multipartite 
entanglement directly from a given density matrix; a general 
method to derive such criteria is missing \cite{partly}.

In this paper we present such a systematic way to develop 
multiparticle entanglement criteria. The resulting criteria 
solve the separability problem for certain families of states  
(notably, the well-studied $N$-qubit GHZ states mixed with white noise) 
and improve known results in many other cases. Also, they allow 
to detect bound entangled states which are separable under 
each partition, but not fully separable. Moreover, our criteria 
can easily be used in todays experiments and  they improve the 
understanding of decoherence in multiparticle quantum systems.

Let us recall the main definitions for multipartite entanglement. 
For three particles, a pure state is fully separable if it is 
of the form $\ket{\psi^{\rm fs}}=\ket{a}\ket{b}\ket{c}$ and a 
mixed state is fully separable if it can be written as a convex 
combination of fully separable pure states
\be
\vr^{\rm fs} = \sum_{k} p_k  \ketbra{\psi^{\rm fs}_k},
\ee
where the $p_k$ form a probability distribution. A pure state 
is called biseparable if it is separable under some 
bipartition. An example is 
$\ket{\psi^{\rm bs}}=\ket{a}\ket{\phi^{bc}}$ where $\ket{\phi^{bc}}$
is a possibly entangled state on the particles $B$ and $C$. This state 
is biseparable under the $A|BC$-partition, other bipartitions 
are the $B|AC$- or $C|AB$-partition. 
A mixed state is biseparable if it can be written as
$
\vr^{\rm bs} = \sum_{k} p_k  \ketbra{\psi^{\rm bs}_k}
$
where the $\ket{\psi^{\rm bs}_k}$ might be biseparable under 
different partitions. Finally a state is genuine multipartite 
entangled, if it is not biseparable. This class of entanglement
one usually aims to generate and verify in experiments 
\cite{genuremark} and we mainly consider entanglement 
criteria for this type of entanglement. 
Note that generalizations and further classifications 
can be found e.g. in Refs.~\cite{seevinckuffink, review, duer99, acin}.

\section{Three qubits} 
We explain our main ideas using three qubits, the 
generalization to more particles (or higher dimensions) 
is straightforward and will be discussed later. 
For a three-qubit density matrix $\vr$ we denote 
its entries by $\vr_{i,j},$ where $1\leq i;j \leq 8,$
here and in the following we always use the standard product basis 
$\{\ket{000},\ket{001},...,\ket{111}\}$. 
Then we have:

{\bf Observation 1.} 
Let $\vr$ be a biseparable three-qubit 
state. Then its matrix entries fulfill
\be
|\vr_{1,8}| \leq 
\sqrt{\vr_{2,2}\vr_{7,7}}+\sqrt{\vr_{3,3}\vr_{6,6}}+
\sqrt{\vr_{4,4}\vr_{5,5}}
\label{eqsobs1}
\ee
and violation implies genuine three-qubit 
entanglement.
\\
{\it Proof.}
First, note that for two positive linear functions  $f(x)$ 
and $g(x)$ the function $ h = \sqrt{fg}$ is concave, that 
is, $h[r x_1+(1-r)x_2] \geq r h(x_1) + (1-r) h (x_2)$ for any 
mixing ratio $r$ \cite{concavityremark}. Consequently, the function 
$\sqrt{\vr_{2,2}\vr_{7,7}}+\sqrt{\vr_{3,3}\vr_{6,6}}+
\sqrt{\vr_{4,4}\vr_{5,5}} - |\vr_{1,8}|$ is concave in the state, 
since it is a sum of concave functions of the matrix entries (the absolute 
value is convex).
So it suffices to 
prove its positivity for pure biseparable states, then 
mixtures of these will inherit the bound. Let 
$
\ket{\psi}= (a_0 \ket{0} + a_1 \ket{1}) \otimes
(b_{00} \ket{00} + b_{01} \ket{01} + b_{10} \ket{10} + b_{11} \ket{11})
$
be a pure state, which is biseparable under the $A|BC$ partition. 
For that, one can directly see that 
$|\vr_{1,8}|=\sqrt{\vr_{4,4}\vr_{5,5}}.$ 
For the other two bipartitions one finds 
$|\vr_{1,8}| = \sqrt{\vr_{3,3}\vr_{6,6}}$ and  
$|\vr_{1,8}| = \sqrt{\vr_{2,2}\vr_{7,7}},$ 
hence, Eq.~(\ref{eqsobs1}) is valid for any pure 
biseparable state, which proves the claim. 
$\qed$

This criterion has also been derived in the context 
of quadratic Bell inequalities \cite{seevinckuffink},
however, our proof is considerably shorter and, most 
importantly, it can be generalized to derive other 
characterizations of the different entanglement classes. 
Note that Eq.~(\ref{eqsobs1}) is independent of the 
normalization of the state, simplifying many calculations below.
Eq.~(\ref{eqsobs1}) is maximally violated by the GHZ state, 
$\ket{GHZ_3}=(\ket{000} + \ket{111})/\sqrt{2}.$ For other 
states, one may first change the local basis (leading, 
e.g., to the criterion 
$|\vr_{2,7}| \leq \sqrt{\vr_{1,1}\vr_{8,8}}+\sqrt{\vr_{3,3}\vr_{6,6}}+
\sqrt{\vr_{4,4}\vr_{5,5}}$), but these will not be considered  
as independent criteria.

To discuss the strength of Observation 1, 
we consider states which are diagonal 
in the GHZ basis. 
This basis consists of the eight states 
$
\ket{\psi_i}=
(\ket{x_1 x_2 x_3} \pm \ket{\bar{x}_1 \bar{x}_2\bar{x}_3})
/\sqrt{2}
$
where $x_j,\bar{x}_j \in \{0 , 1\}$ and $x_j \neq \bar{x}_j.$
States which are diagonal in this basis are of the form
\be
\vr^{\rm(dia)}=
\frac{1}{\NN}
\begin{bmatrix}
\lambda_1 & 0 & 0 & 0 & 0 & 0 & 0 & \mu_1
\\
0 &\lambda_2 &  0 & 0 & 0 & 0  & \mu_2 &0
\\
0 &0 &\lambda_3  & 0 & 0  & \mu_3 &0&0
\\
0 & 0 &0 &\lambda_4  &  \mu_4 &0&0&0
\\
0 & 0 &0 &\mu_4  &  \lambda_5 &0&0&0
\\
0 &0 &\mu_3  & 0 & 0  & \lambda_6 &0&0
\\
0 &\mu_2 &  0 & 0 & 0 & 0  & \lambda_7 &0
\\
\mu_1 & 0 & 0 & 0 & 0 & 0 & 0 & \lambda_8
\end{bmatrix}
\label{zustandsfamilie}
\ee
with real  $\lambda_i$ and  $\mu_i$, fulfilling
$\lambda_i=\lambda_{9-i}$ for $i=1,...,4$, and
$\NN$ denotes a normalization. 
We can state:

{\bf Observation 2.} For GHZ-diagonal states, the criterion
from Observation 1 constitutes a necessary and sufficient 
criterion for genuine multipartite entanglement. 
\\
{\it Proof.} The proof is given in the Appendix.
$\qed$

This shows that the criterion of Observation 1 is a strong 
criterion in the vicinity of GHZ states, indeed its later 
generalization solves the problem of classifying $N$-qubit 
GHZ states mixed with white noise (see Fig. 1). 

It remains to investigate what happens for other states, 
such as the W state,
$\ket{W_3}=(\ket{001}+ \ket{010} + \ket{100})/\sqrt{3}.$
First, one can apply local unitary operations, before testing 
Eq.~(\ref{eqsobs1}). This indeed works for the pure W state, 
but one can also derive stronger criteria:

{\bf Observation 3.} Any biseparable three-qubit state 
fulfills
\bea
|\vr_{2,3}| + |\vr_{2,5}| + |\vr_{3,5}|
&\!\leq\!&\!
\sqrt{\vr_{1,1}\vr_{4,4}}
+\sqrt{\vr_{1,1}\vr_{6,6}}+
\sqrt{\vr_{1,1}\vr_{7,7}}
\nonumber
\\
&& + \frac{1}{2}(\vr_{2,2}+\vr_{3,3}+\vr_{5,5}).
\label{eqsobs3}
\eea
{\it Proof.} Again, it suffices to consider pure states. 
Then, for a state which is $A|BC$-biseparable one sees that
$|\vr_{2,5}|=\sqrt{\vr_{1,1}\vr_{6,6}}$
and
$|\vr_{3,5}|=\sqrt{\vr_{1,1}\vr_{7,7}}.$ 
Furthermore, one has 
$|\vr_{2,3}| \leq (\vr_{2,2}+\vr_{3,3})/2$
which follows already from the positivity of 
the density matrix. 
Therefore, Eq.~(\ref{eqsobs3}) holds 
for the $A|BC$ partition, and similarly one can 
prove  it holds for the other two bipartitions. 
$\qed$

This observation deserves two comments. First, this 
criterion is quite strong: It detects W states mixed 
with white noise, i.e., 
$\vr^{\rm (w3)}(p) = (1-p)\ketbra{W_3} + p {\eins}/{8},$
for $p < 8/17 \approx 0.471$ as genuine tripartite entangled, 
while the best known entanglement witness detects it only for 
$p < 8/19 \approx 0.421$ \cite{witremark}.
Second, it should be noted that Observation 3 is independent 
of Observation 1: The states $\vr^{\rm (w3)}(p)$ for 
$p \in (0.413; 0.471)$ are directly detected by Observation 3. 
They are, however, not detected by Eq.~(\ref{eqsobs1}), even 
if local filter operations 
$\vr \mapsto \tilde{\vr}
= \FF_1 \otimes \FF_2 \otimes \FF_3 \vr 
\FF_1^\dagger \otimes \FF_2^\dagger \otimes \FF_3^\dagger$
are applied with arbitrary matrices $\FF_i,$ as can be checked numerically.

\begin{figure}[t]
\includegraphics[scale=1.1]{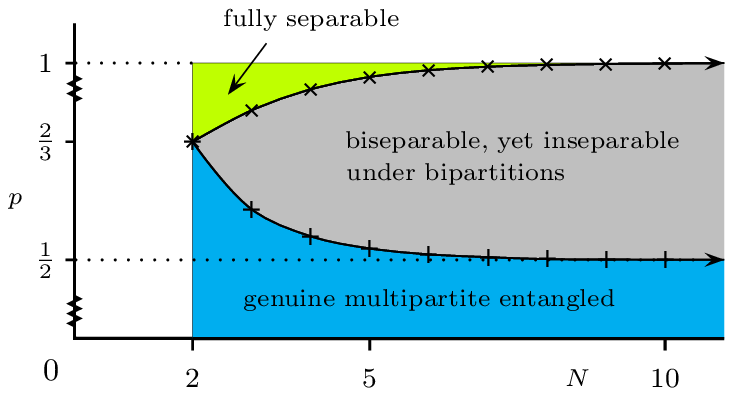}
\caption{(color online) The entanglement 
properties of $N$-qubit GHZ states mixed with white noise, 
$\vr^{\rm (ghzN)}=(1-p)\ketbra{GHZ_N} + p \eins /2^N$.
It was known before \cite{duer99} that these are 
fully separable iff $1/[1+2^{(1-N)}] \leq p \leq 1$, 
while for smaller $p$ they are inseparable under any 
partition. Our results show, that iff $0\leq p <1/[2(1-2^{-N})]$ 
the states are genuinely multipartite entangled. Consequently, 
in the region in between the two bounds the states 
$\vr^{\rm (ghzN)}$ are biseparable yet inseparable 
under any  fixed bipartition. 
} 
\label{figGHZ}
\end{figure}

So far, we have only considered criteria for biseparable 
states. Our approach also allows to derive criteria for 
other entanglement classes:

{\bf Observation 4.} 
(i) For fully separable three-qubit states, the following
 inequalities hold:
\bea
|\vr_{1,8}| &\leq & 
\big(\vr_{2,2}\cdot \vr_{3,3} \cdot \vr_{4,4}\cdot \vr_{5,5} \cdot \vr_{6,6}
\cdot \vr_{7,7}\big)^{\frac{1}{6}}
\label{obs4eq1}
\\
|\vr_{1,8}| &\leq & 
\big(\vr_{1,1}\cdot \vr_{4,4}^2\cdot \vr_{5,5} \cdot \vr_{6,6}
\cdot \vr_{7,7}\big)^{\frac{1}{6}}
\label{obs4eq2}
\eea
(ii) Eq.~(\ref{obs4eq1}) and Eq.~(\ref{obs4eq2}) are 
connected via the substitution 
$\vr_{2,2}\vr_{3,3} \rightarrow \vr_{1,1}\vr_{4,4}.$ 
Similarly, one obtains new separability criteria from 
Eq.~(\ref{obs4eq1}) by making the substitutions
$\vr_{6,6}\vr_{7,7} \rightarrow \vr_{5,5}\vr_{8,8},$ 
$\vr_{2,2}\vr_{5,5} \rightarrow \vr_{1,1}\vr_{6,6},$ 
$\vr_{4,4}\vr_{7,7} \rightarrow \vr_{3,3}\vr_{8,8},$ 
$\vr_{3,3}\vr_{5,5} \rightarrow \vr_{1,1}\vr_{7,7},$ 
and
$\vr_{4,4} \vr_{6,6} \rightarrow \vr_{2,2}\vr_{8,8}.$ 
Combining such substitutions, one  also obtains new 
separability criteria, 
e.g.
$
|\vr_{1,8}| \leq  
(\vr_{2,2} \cdot \vr_{3,3} \cdot \vr_{5,5} \cdot \vr_{8,8})^{{1}/{4}}.$
\\
(iii) A condition for full separability which is violated
in the vicinity of a W state is
\be
|\vr_{2,3}| + |\vr_{2,5}| + |\vr_{3,5}|
\leq
\sqrt{\vr_{1,1}\vr_{4,4}}+\sqrt{\vr_{1,1}\vr_{6,6}}
+\sqrt{\vr_{1,1}\vr_{7,7}}.
\label{obs4eq3}
\ee
(iv) Eq.~(\ref{obs4eq1}) is a necessary and 
sufficient criterion for full separability for GHZ states 
mixed with white noise.
\\
{\it Proof.} The proof is essentially the same as before, 
using the concavity of more generalized functions \cite{concavityremark}. 
The inequalities (\ref{obs4eq1},\ref{obs4eq3}) are equalities for pure 
fully separable states. The substitutions as
in Eq.~(\ref{obs4eq2}) can be made, since 
$\vr_{2,2}\vr_{3,3}=\vr_{1,1}\vr_{4,4}$, etc.~holds 
for any pure fully separable state. Concerning (iv), 
note that  Eq.~(\ref{obs4eq1}) detects noisy GHZ 
states for $p<4/5,$ and this value is known to mark the
border of the fully separable 
states \cite{duer99}. 
$\qed$

Surprisingly, substitutions  as in Eq.~(\ref{obs4eq2})
do indeed improve the criterion in some cases. For example, consider the 
family of bound entangled states of Ref.~\cite{acin}. These are states as in 
Eq.~(\ref{zustandsfamilie}) 
with 
$\lambda_1=\lambda_8=\mu_1=1$
and 
$\lambda_2=1/\lambda_7,$
$\lambda_3=1/\lambda_6,$
$\lambda_4=1/\lambda_5,$
and $\mu_2=\mu_3=\mu_4=0.$
For $\lambda_2 \cdot \lambda_3 \neq \lambda_4$ 
these states are separable under each bipartition, 
but not fully separable. Their entanglement is  
detected by Eq.~(\ref{obs4eq2}) or other 
substitutions. Moreover, as one can directly check, 
for the special case $\lambda_2=\lambda_3=\lambda_5$
the inequality in (ii) tolerates significantly more 
noise than the best known witness \cite{hyllus} and 
gives more significant results for recent experiments.
\cite{hermannremark}.

\section{Many qubits}
Let us start with introducing a compact notation.
First, we label the diagonal elements of $\vr$ 
by the corresponding product vector 
in the standard basis. That is, if $I=(i_1, i_2, ..., i_N)$
is a tuple consisting of $N$ indices $i_k \in \{0,1\}$
then $\vr_I = \vr_{(i_1, i_2, ..., i_N)}$ is the diagonal
entry corresponding to $\ketbra{i_1, i_2, ..., i_N}.$ For 
example, for three qubits $\vr_{(000)}=\vr_{1,1}$
and  $\vr_{(001)}=\vr_{2,2}$ etc. For a given $I$ one can 
define $\overline{I}$ as the tuple arising from $I$ if zeroes 
and ones are exchanged, e.g., $\overline{(001)}=(110)$
Furthermore, let $|I|$ denote the number of $i_k = 1$ in 
$I,$ then $\sum_{|I|=n}$ denotes a sum over all $I$ with $|I|=n.$

Second, let $\sigma = \ketbra{\psi}$ be a target state 
and  $\vr$ be a different state. We abbreviate with 
${\fO}^{\ket{\psi}}(\vr)$ the sum of the absolute values 
of the off-diagonal elements of $\vr$ in the upper triangle, 
which correspond to matrix entries where $\sigma$ does not vanish.
For instance, for the three-qubit GHZ state
we have ${\fO}^{\ket{GHZ_3}}(\vr) = |\vr_{1,8}|$ and 
Eq.~(\ref{eqsobs3}) can now be conveniently rewritten as
${\fO}^{\ket{W_3}}(\vr) \leq 
\sum_{|I|=2}\sqrt{\vr_{(000)}\vr_I}+
\tfrac{1}{2}\sum_{|I|=1}\vr_I$. 

The idea behind this notation is to estimate all off-diagonal 
elements similarly as in Observation 1. Explicitely, we have for
four qubits:

{\bf Observation 5.} 
(i) From the four-qubit GHZ state, 
$\ket{GHZ_4} = (\ket{0000} + \ket{1111})/\sqrt{2},$
a necessary condition for biseparability  of a general 
state $\vr$ is
\be
{\fO}^{\ket{GHZ_4}}(\vr) 
\leq 
\frac{1}{2}\sum_{|I|=\{1,2,3\}}
\sqrt{\vr_{I \vphantom{\overline{I}}}\vr_{\overline{I}}}.
\label{obs5N}
\ee
This condition is necessary and sufficient for biseparability of 
GHZ-diagonal states in the sense of Observation 2.
\\
(ii) From the four-qubit W state, 
$\ket{W_4} = (\ket{0001} + \ket{0010}+\ket{0100} + \ket{1000}) /{2},$
a criterion is derived as
\be
{\fO}^{\ket{W_4}}(\vr) \leq 
\sum_{|I|=2}\sqrt{\vr_{(0000)}\vr_I}+
\sum_{|I|=1}\vr_I.
\label{obs5eqs2}
\ee
(iii) From the four-qubit Dicke state, 
$\ket{D_4} = 
(\ket{0011} + \ket{0101} + \ket{1001}
+\ket{0110}+\ket{1010}+\ket{1100}) /\sqrt{6}$,
a criterion is derived as
\be
{\fO}^{\ket{D_4}}(\vr) \leq 
\sqrt{\vr_{(0000)}\vr_{(1111)}}
+
\sum_{|I|=1}\sum_{|J|=3}\sqrt{\vr_{I \vphantom{J}}\vr_{{J}}}+
\frac{3}{2}
\sum_{|I|=2}\vr_I.
\label{obs5eqs3}
\ee
{\it Proof.} (i) is proved as in Observations 1 and 2. 
The factor $1/2$ takes into account that each possible 
term occurs twice in the sum. (ii) and (iii) follow 
as in Observation 3. Here, estimating an
off-diagonal element can be simplified 
by the following rule: If the off-diagonal 
element $\eta$ corresponds to 
$\ket{i_1 i_2 i_3 i_4}\bra{j_1 j_2 j_3 j_4}$ 
and the state is separable under the $A|BCD$-bipartition, 
one has 
$\eta \leq \sqrt{\vr_{(i_1 j_2 j_3 j_4)} \vr_{(j_1 i_2 i_3 i_4)}}$
while one has 
$\eta \leq \sqrt{\vr_{(i_1 i_2 j_3 j_4)} \vr_{(j_1 j_2 i_3 i_4)}}$
for the $AB|CD$-bipartition, etc. Further, one needs that for a 
positive $n \times n$ matrix $P$ the bound 
$\sum_{i <  j}|P_{ij}| \leq \tfrac{n-1}{2} Tr(P)$ holds \cite{positivityremark}.
$\qed$

Again, these criteria improve known conditions: For the four-qubit W 
state mixed  with white noise, Eq.~(\ref{obs5eqs2}) detects genuine 
multipartite entanglement for $p < 4/9 \approx 0.444,$ while the 
fidelity based witness detects it only for $p < 4/15 \approx 0.267$ 
and the improved witness \cite{witremark} for  $p < 16/45 \approx 0.356.$ 
A four-qubit Dicke state mixed with white noise is detected by 
Eq.~(\ref{obs5eqs3}) for $p < 8/21 \approx 0.381,$ while the best 
known witness detects it for $p > 16/45 \approx 0.356$ 
\cite{gezadicke}.

For arbitrary states similar entanglement criteria can be derived as 
follows: In a given basis and for a fixed partition, any off-diagonal 
element can be estimated as in the proof of Observation 5. Then, all 
these estimates can be summarized to an estimate of the sum of all 
off-diagonal elements. This might be further improved by considering 
a weighted sum. For instance, for $N$-qubit GHZ states, the 
criterion reads 
$
{\fO}^{\ket{GHZ_N}}(\vr) 
\leq 
\tfrac{1}{2}
\sum_{|I|=1}^{N-1} \sqrt{\vr_{I \vphantom{\overline{I}}}\vr_{\overline{I}}},
$
and is again necessary and sufficient for GHZ diagonal states as the proof 
of Observation 2 can directly be generalized (see Fig.~1). Further criteria 
for  cluster states or the four-qubit singlet state will be presented 
elsewhere.

\section{Experimental consequences} 
Obviously, these criteria can be applied to experiments  
where the full density matrix has been determined 
\cite{haeffner}. However, often this can not be done. 
Still, our results may be directly applied. For 
example, let us consider Eq.~(\ref{eqsobs3}) for the 
detection of entanglement around the three-qubit W state.
Using the fidelity $F = Tr(\vr \ketbra{W_3})$ one may 
rewrite Eq.~(\ref{eqsobs3}) as
\be
F
\leq
\frac{2}{3}
\big(
\sqrt{\vr_{1,1}\vr_{4,4}}
+\sqrt{\vr_{1,1}\vr_{6,6}}+
\sqrt{\vr_{1,1}\vr_{7,7}}
+\vr_{2,2}+\vr_{3,3}+\vr_{5,5}
\big).
\nonumber
\ee
The fidelity of the W state can be measured experimentally 
with five local measurements \cite{ijtp} and the diagonal 
elements can also be determined from measurement of 
$\sigma_z \otimes \sigma_z \otimes \sigma_z,$ which is already 
included in the measurements needed for the fidelity. 
This shows that Eq.~(\ref{eqsobs3}) (and similarly all 
other criteria presented) is experimentally easily 
testable. For the usual error models in photon 
experiments one can also check that criterion 
(\ref{eqsobs3}) detects entanglement with a 
higher statistical significance than the witness, 
unless the fidelity is close to one and the significance 
of both methods is high.

\section{Decoherence} 
Finally, our results also shed light on the decoherence 
of multipartite entanglement. Consider an N-qubit GHZ 
state, influenced by relaxation -- the noise that is 
dominant in ion traps \cite{roos}. On a single qubit, 
this changes the density matrix according to 
$\ketbra{0} \rightarrow \ketbra{0}$,
$\ketbra{1} \rightarrow x \ketbra{1} + (1-x) \ketbra{0}$,
and 
$(\ket{0}\bra{1} + h.c.)  \rightarrow x^{1/2} (\ket{0}\bra{1}+ h.c.)$
(with $x=e^{-\gamma t}$) and corresponds to a coupling to 
a bath with zero temperature. 
The total density matrix can directly be 
computed \cite{aolita}, resulting in $\vr_{1,2^N}=x^{N/2}$  for 
the off-diagonal element and 
$\vr_I = [\delta_{|I|,0}+x^{|I|}(1-x)^{N-|I|}]/2$ for the diagonal 
elements. Here, we have used the same notation as in Observation 5.

This state is not diagonal in the GHZ basis, but applying on each
qubit a filter $\vr \mapsto \FF \vr \FF$ with $\FF = \alpha \ketbra{0} 
+ (1/\alpha)\ketbra{1}$ and $\alpha^4=x/(1-x)$ maps it to a state that 
differs from a GHZ diagonal state only in the element $\vr_{1,1}.$
This filtering keeps all entanglement properties, but finally 
Observations 1 and 2 can be used. From this one can conclude that
GHZ states coupled to a bath with zero temperature are genuine 
multipartite entangled, 
if and only if $t < - \ln[1-(2^{N-1}-1)^{-2/N}]/\gamma$.

\section{Conclusion}
We presented a method to derive separability 
criteria for different classes of multipartite 
entanglement directly in terms of density matrix 
elements. The resulting criteria are strong and 
can be used in experiments, as well as for the 
investigation of decoherence. 
It would be interesting to use our approach 
to discriminate between more special entanglement classes (such 
as the W and GHZ class for three qubits \cite{acin}) 
and to connect it to the quantification of entanglement 
with entanglement measures.

We thank J. Uffink for fruitful discussions.
This work has been supported by the FWF (START prize) 
and the EU (OLAQUI, QICS, SCALA). MPS acknowledges 
the hospitality of the Centre for Time, University of Sydney.

{\it Note added:}
Half a year after submission of our manuscript to the arxiv, a preprint 
\cite{huber} has appeared, in which criteria for multipartite entanglement 
have been presented. This method is analog to ours, by using estimates 
for off-diagonal terms [cf.~Eq.~(I) in Ref.~\cite{huber} with the proof 
of our Observation 5] and using convexity arguments. Consequently, the 
obtained separability criterion used for N-qubit GHZ states [Eq.~(II) in 
Ref.~\cite{huber}] is the same as our criterion in Eqs.~(\ref{eqsobs1}, \ref{obs5N})
combined with local filtering operations,
and the criterion 
for W states  [Eq. (III) in Ref.~\cite{huber}] is for three 
qubits the same as our Eq.~(\ref{eqsobs3}), while for four qubits 
it is weaker than our Eq.~(\ref{obs5eqs2}).

\section*{Appendix} 
Here, we prove Observation 2. 
Since $\vr^{\rm(dia)} \geq 0$
we have  $|\mu_i| \leq \lambda_i$
and we can assume that 
$\lambda_1 \geq \lambda_2 \geq \lambda_3 \geq \lambda_4$ 
as one can achieve that by a local change of the 
basis. Then, Eq.~(\ref{eqsobs1}) reads  
$|\mu_1| \leq \lambda_2 + \lambda_3 + \lambda_4,$ and we 
will show that if this holds, a decomposition
into biseparable states can be found. 
Note that due to the ordering of the $\lambda_i$ 
other conditions for biseparability (e.g. 
$|\mu_2| \leq \lambda_1 + \lambda_3 + \lambda_4$)
can then never be violated.

Let us define the unnormalized state $\vr^{(12)}(\lambda)$ 
with $\lambda_1 = \mu_1=\lambda_2 = \mu_2 = \lambda$, 
while all other matrix entries vanish. This state is 
$AB|C$-biseparable, since it can be written as
\begin{align}
\vr^{(12)}(\lambda) = &  
2 \lambda
\big(
\ketbra{\chi^+}_{AB} \otimes \ketbra {\eta^+}_{C}+
\nonumber
\\
&+  
\ketbra{\chi^-}_{AB} \otimes \ketbra {\eta^-}_{C}
\big)
\end{align}
with $\ket{\chi^\pm}= (\ket{00} \pm \ket{11})/\sqrt{2}$
and $\ket{\eta^\pm}= (\ket{0} \pm \ket{1})/\sqrt{2}.$
Analogously, one can consider states 
$\vr^{(kl)}$ for any $k,l = 1,...,4$ 
with $k \neq l$ and finds that they are 
also biseparable, as one only has to permute or flip some qubits.

(i) First, we consider the extremal case when 
$\mu_i = \lambda_i$ for all $i$ and by assumption 
the separability condition implies that we have 
$\mu_i = \lambda_i \leq \sum_{k \neq i}\lambda_k,$
where for the index $1 \leq k \leq 4.$
If $\lambda_1 = \lambda_2 + \lambda_3 + \lambda_4$
we can directly write 
$
\vr^{\rm(dia)} = \sum_{k=2,3,4} 
\vr^{(1k)}(\lambda_k)
$
hence $\vr^{\rm(dia)}$ is biseparable.  
Otherwise, the idea is to write 
\be
\vr^{\rm(dia)} 
= 
\sum_{k=2,3,4} 
\vr^{(1k)}(\chi_k)
+\vr^{\rm (r)}
\ee
for some parameters $\chi_k$ such that the rest $\vr^{\rm (r)}$ 
(which is then characterized by parameters $\lambda^{\rm (r)}_k$)
fulfills two conditions. Its first and last column and row 
should vanish ($\vr^{\rm (r)}_{1,1}= \lambda^{\rm (r)}_1=0$) 
and it should still fulfill all biseparability conditions 
(e.g. $\lambda^{\rm (r)}_2 \leq \lambda^{\rm (r)}_3+\lambda^{\rm (r)}_4$). 
Then, $\vr^{\rm (r)}$ can be iteratively further decomposed and 
finally a decomposition of $\vr^{\rm(dia)}$ into biseparable states
can be found.

The idea is to choose the $\lambda^{\rm (r)}_k, k=2,3,4$ 
as equal as possible (they have to fulfill 
$\lambda^{\rm (r)}_k \leq \lambda_k$), 
but monotonically decreasing. For that, we define 
$\alpha_4 
:= \lambda_2 + \lambda_3 + \lambda_4 -\lambda_1 
= \lambda^{\rm (r)}_2 + \lambda^{\rm (r)}_3+\lambda^{\rm (r)}_4 > 0$
and then recursively
$\lambda^{\rm (r)}_4= \min\{\lambda_4, \alpha_4/3\}$, 
then $\alpha_3= \alpha_4 - \lambda^{\rm (r)}_4$ and then 
$\lambda^{\rm (r)}_3 = \min\{ \lambda_3, \alpha_3/2\}$ and 
finally $\alpha_2= \alpha_3 - \lambda^{\rm (r)}_3$ and 
$\lambda^{\rm (r)}_2 = \min\{ \lambda_2, \alpha_2\}.$
Then 
$\vr^{\rm(dia)} 
= 
\sum_{k=2,3,4} 
\vr^{(1k)}(\lambda_k - \lambda^{\rm (r)}_k)
+\vr^{\rm (r)}
$
with 
$\lambda^{\rm (r)}_2 \geq \lambda^{\rm (r)}_3 \geq \lambda^{\rm (r)}_4.$
Then we cannot have that both
$\lambda^{\rm (r)}_4=\lambda_4$
and 
$\lambda^{\rm (r)}_3=\lambda_3,$
because if this were true, then from the definition 
of $\alpha_4$ it would follow that $\lambda^{\rm (r)}_2 = \lambda_ 2 - \lambda_1 \leq 0$. 
So we have $\lambda^{\rm (r)}_3=\alpha_3/2$ 
($\lambda^{\rm (r)}_4=\alpha_4/3$ also
implies $\lambda^{\rm (r)}_3=\alpha_3/2$),
which due to the ordering of the $\lambda_i$
implies $\lambda^{\rm (r)}_2=\alpha_2 = \lambda^{\rm (r)}_3 $.
So $\lambda^{\rm (r)}_2 \leq \lambda^{\rm (r)}_3+\lambda^{\rm (r)}_4$
and one can decompose $\vr^{\rm (r)}$ further into $\vr^{(23)}$ 
and $\vr^{(24)}$ and a remaining term with 
${\lambda}^{\rm (rr)}_1 = {\lambda}^{\rm (rr)}_2 =0$ etc.
Of course, for the case of three qubits one may also write 
down suitable values for the $\lambda^{\rm (r)}_i$ directly, 
but the previous scheme can directly be extended to more qubits.

(ii)  Secondly, for $0 \leq \mu_i \leq \lambda_i$,  and 
where again $\lambda_1 \leq  \lambda_2 + \lambda_3 + \lambda_4$, 
we first consider the states $\vr^{(kl)}$. Their non-zero 
matrix elements obey ${\mu}_i = {\lambda}_i$, 
but, applying with some probability locally conjugate random 
phases (e.g, $\ket{1}_2 \mapsto e^{i \phi}\ket{1}_2$ and 
$\ket{1}_3 \mapsto e^{-i\phi}\ket{1}_3$) to these states 
decreases the values of the ${\mu}_i$ (in this 
example for $i=2,3$). Therefore, one can for a given 
$\vr^{(kl)}$ decrease the values of ${\mu}_i$ arbitrarily
by local operations (in the example we can decrease e.g. the value of 
${\mu}_2$ for $\vr^{(12)}$ or ${\mu}_3$ for $\vr^{(34)}$), 
and the resulting states must be 
biseparable. Consequently, a given $\vr^{\rm(dia)}$ with 
$\lambda_1 \leq  \lambda_2 + \lambda_3 + \lambda_4$
can be decomposed into biseparable states as in (i).

(iii) Further, it may happen that for a given $\vr^{\rm(dia)}$
one has  $0 \leq \mu_1 \leq  \lambda_2 + \lambda_3 + \lambda_4$
but
$\lambda_1 >  \lambda_2 + \lambda_3 + \lambda_4.$
Then we consider  $\hat{\vr}$ which is obtained from $\vr^{\rm(dia)}$
by setting $\lambda_1 = \max\{\mu_1, \lambda_2\}.$ Then, $\hat{\vr}$ 
is biseparable according to (ii), and $\vr^{\rm(dia)}$ is obtained from
$\hat{\vr}$ by mixing with the fully separable state 
$\ketbra{000}+\ketbra{111},$ hence it is biseparable. 

(iv) The previous arguments prove the claim 
if all $\mu_i \geq 0.$ If some $\mu_i$ are negative, 
one can prove it as follows: Let $\vr$ be a GHZ diagonal 
state, with some $\mu_i < 0$, which fulfills the condition 
of biseparability. The state $\widehat{\vr}$ which arises from $\vr$ 
when all $\mu_i$ are replaced by $|\mu_i|$ fulfills the same condition, 
and is biseparable due to points (i)-(iii). It can be decomposed into 
several $\vr^{(kl)}$, maybe in some of them we have 
$\mu_i(\vr^{(kl)}) < \lambda_i(\vr^{(kl)})$ according to points (ii) and 
(iii). Nevertheless, we can built out of this decomposition of 
$\widehat{\vr}$ a decomposition of $\vr$, if we flip the signs of all the 
$\mu_i(\vr^{(kl)})$ appropriately. An arbitrary flipping of the signs of 
the $\mu_i$  of a given $\vr^{(kl)}$ can be done for each $k,l$ by local 
operations, hence $\vr$ is also biseparable.
$\qed$



\begin{thebibliography}{99}
\bibitem{concavityremark} 
More generally, one has that if $f_1, ... , f_n$
are positive concave functions, then 
$g=\big(\prod_{k=1}^{n}f_k \big)^{1/n}$
is also concave. This can be seen as follows: 
First, as the function $h(x)=(x)^{1/n}$ is monotonically 
increasing, it suffices to prove the claim for linear 
$f_k.$ Then, one can directly calculate that the second 
derivative of $g$ is not positive. See also p. 87 in
S. Boyd and L. Vandenberghe, {\it Convex optimization}, 
(Cambridge University Press, 2004).
\bibitem{witremark} 
For three qubits, this 
witness is $\WW = (2/3)\cdot (\eins -\ketbra{111}) - \ketbra{W_3},$
see also Section 6.8.2 in Ref.~\cite{review}.
\bibitem{positivityremark}
This generalizes the estimate 
$|\vr_{23}| \leq (\vr_{22}+\vr_{33})/2$
from Observation 3 and can be seen as follows: 
One has for any $\ket{x}$ that $\bra{x}P\ket{x} \geq 0,$
and taking $\ket{x}$ of the type 
$\ket{x}=(1,e^{i \phi},0,...0)$
and summing over all possible 
permutations thereof gives the bound.
\end{thebibliography}
\end{document}